\documentclass[fleqn,usenatbib]{mnras}

\usepackage{newtxtext,newtxmath}

\usepackage[T1]{fontenc}

\DeclareRobustCommand{\VAN}[3]{#2}
\let\VANthebibliography\thebibliography
\def\thebibliography{\DeclareRobustCommand{\VAN}[3]{##3}\VANthebibliography}

\usepackage{graphicx}	
\usepackage{amsmath}	
\usepackage{amssymb}	
\usepackage{multirow}
\usepackage{threeparttable}

\title[Magnetized reverse shock forming condition]{Reverse shock forming condition for magnetized relativistic outflows: reconciling theories and simulations}

\author[J.-Z. Ma and B. Zhang]{
Jing-Ze Ma$^{1}$\thanks{E-mail: mjz18@mails.tsinghua.edu.cn}
and Bing Zhang$^{2,3}$\thanks{E-mail: zhang@physics.unlv.edu}
\\
$^{1}$School of Aerospace Engineering, Tsinghua University,
 Beijing 100084, China\\
$^{2}$Nevada Center for Astrophysics, University of Nevada Las Vegas, Las Vegas, NV 89154, USA\\
$^{3}$Department of Physics and Astronomy, University of Nevada Las Vegas, Las Vegas, NV 89154, USA
}

\date{Accepted XXX. Received YYY; in original form ZZZ}

\pubyear{2015}

\begin{document}
\label{firstpage}
\pagerange{\pageref{firstpage}--\pageref{lastpage}}
\maketitle

\begin{abstract}
Reverse shock (RS) emission can be used to probe the properties of the relativistic ejecta, especially the degree of magnetization $\sigma$, in gamma-ray burst (GRB) afterglows.
However, there has been confusion in the literature regarding the physical condition for the RS formation, and the role of magnetic fields in the RS dynamics in the Poynting-flux-dominated regime is not fully understood. 
Exploiting the shock jump conditions, we characterize the properties of a magnetized RS.
We compare the RS dynamics and forming conditions from different theories and numerical simulations, and reconcile the discrepancies among them.
The strict RS forming condition is found to be $\sigma < \sigma_\mathrm{cr}=(8/3)\gamma_4^2(n_1/n_4)$, where $n_4$ and $n_1$ are the rest-frame number densities of the ejecta and the ambient medium, respectively, $\gamma_4$ is the bulk Lorentz factor, and $\sigma_\mathrm{cr}$ is the critical magnetization.
Contrary to previous claims, we prove that this condition agrees with other theoretical and simulated results, which can be further applied to the setup and consistency check of future numerical simulations.
Using this condition, we propose a characteristic radius for RS formation, 
and categorize the magnetized shell into three regimes: `thick shell' (relativistic RS), `thin shell' (trans-relativistic RS), and `no RS' regimes.
The critical magnetization $\sigma_\mathrm{cr}$ is generally below unity for thin shells, but can potentially reaches $\sim 100-1000$ in the `thick shell' regime.
Our results could be applied to the dynamical evolution of Poynting-flux-dominated ejecta, with potential applications to self-consistent lightcurve modelling of magnetized relativistic outflows.

\end{abstract}

\begin{keywords}
MHD -- shock waves -- gamma-ray burst: general -- ISM: jets and outflows -- galaxies: jets
\end{keywords}



\section{Introduction}

Relativistic shock waves often develop when a relativistic ejecta is decelerated by the ambient medium.
Under certain circumstances, the excited waves would steepen into a reverse shock (RS) propagating into the ejecta as well as a forward shock (FS) propagating into the medium.
Relativistic shocks are believed to be efficient particle accelerators. The standard FS-RS model is widely invoked to interpret the nonthermal emission in various astrophysical systems, such as pulsar wind nebulae \citep[e.g.][]{kennelConfinement1984A}, gamma-ray bursts \citep[GRBs, e.g.][]{reesRelativistic1992M}, and fast radio bursts \citep[FRBs, e.g.][]{lyubarskymodel2014M}.

In particular, the RS radiation signature in GRB afterglows can serve as a tool to probe the properties of the ejecta \citep[for a review of GRBs and the role of RS, see][]{kumarphysics2015PRb}.
Both the RS and FS emissions may contribute to the GRB early afterglows.
The composition and the bulk Lorentz factor of the ejecta can be constrained by analyzing the afterglow light curves, e.g., from the presence (or absence) of the RS emission and the relative importance of RS and FS components \citep{sariGRB1999A, sariPredictions1999A, kobayashiLight2000A, kobayashiGRB2003A, zhangGammaRay2003A, nakarEarly2004M, harrisonMagnetization2013A, japeljPhenomenology2014A, gaoMorphological2015A}.

One of the ejecta properties reflected by the RS emission is the degree of magnetization $\sigma$, the ratio of the Poynting flux to the matter energy flux measured in the lab frame.
When the magnetic field is dynamically unimportant ($\sigma \ll 1$), the RS is classified into two regimes: the `thick shell' case where a relativistic RS (RRS) decelerates the ejecta, and the `thin shell' case where the RS is Newtonian \citep{sariHydrodynamic1995A, kobayashiLight2000A}.
The relative magnetization of the RS with respect to the FS can be deduced from the afterglow light curves, thereby indirectly imposing constraints on the magnetization of the ejecta \citep{zhangGammaRay2003A, harrisonMagnetization2013A, japeljPhenomenology2014A, gaoMorphological2015A, huangVery2016A, jordana-mitjansLowly2020A}.

However, the physics of the RS remains not well-understood in the high-$\sigma$ regime.
The general picture is as follows: As $\sigma$ increases from below, the RS emission becomes stronger due to the progressively stronger synchrotron radiation until reaching a peak at $\sigma \sim 0.1-1$, and then weakens as the RS becomes Newtonian and suppressed by a dynamically-important magnetic field \citep{fanvery2004A, zhangGamma2005Aa, chenenergyconserving2021M}.
This is partly supported by the particle acceleration mechanism.
For ultra-relativistic shocks, numerical simulations suggested that particles can only be accelerated efficiently via the diffusive shock acceleration when $\sigma \lesssim 10^{-3}$ \citep{sironiParticle2009A, sironiParticle2011A, sironiMaximum2013A}, but precursor waves generated by the synchrotron maser instability may operate as an alternative acceleration mechanism up to $\sigma \sim 0.1-1$ \citep{gallantRelativistic1992A, iwamotoPersistence2017A, iwamotoPrecursor2018A, plotnikovsynchrotron2019M, sironiCoherent2021PRL, iwamotoParticle2021a}.
It is less clear whether these conclusions hold for mildly relativistic shocks \citep{sironiParticle2011A, ligoriniMildly2021M}, which could be important for a trans-relativistic RS.
The allowed range of magnetization for a detectable RS emission is thus still poorly constrained, both from theoretical and observational perspectives.

A primary step to understand this problem is to study the $\sigma$-dependent forming condition of the RS, which is under debate in the literature.
In a one-dimensional (1D) Cartesian geometry, \citet{zhangGamma2005Aa} (hereafter ZK05) found analytically that the RS cannot form when the magnetization is higher than a critical value $\sigma_\mathrm{ cr} = (8/3)\gamma_4^2 ({n_1}/{n_4})$, where $n_4$ and $n_1$ are the rest-frame number densities of the ejecta and the ambient medium, respectively, and $\gamma_4$ is the bulk Lorentz factor ($\sigma_\mathrm{ cr} \sim 100$ for typical GRB conditions).
However, it was argued that although this criterion is correct in 1D magnetohydrodynamic (MHD) Riemann problem \citep[as supported by the numerical simulations of][]{2009ApJ...690L..47M}, it may not hold in reality due to the multi-dimensional instability of shocks \citep{lyutikovSimple2010PRE, lyutikovDynamics2011M} or density fluctuations \citep{granotInteraction2012M}.
Alternatively, \citet{gianniosexistence2008A} (hereafter GMA08) claimed that the ejecta may adjust its structure through fast magnetosonic waves before the RS crosses the ejecta, thus giving a tighter constraint ($\sigma_\mathrm{ cr} \sim 1$ for typical GRB conditions).
This argument was later criticized by \citet{granotInteraction2012M} because it contradicts with the fact that a fast MHD shock always travels faster than any linear wave in the upstream.
Detailed numerical simulations by \citet{mimicaDeceleration2009A} (hereafter MGA09) found near-Newtonian RSs, though with low dissipative efficiencies, in the ``no RS'' regime claimed by GMA08.
However, MGA09 showed that the near-Newtonian RS emerges from a rarefaction fan instead of the contact discontinuity (CD) due to the evolution of the flow in spherical geometry, which cannot be captured by the 1D Cartesian geometry as adopted in ZK05 and \citet{2009ApJ...690L..47M}.
The confusion persists even to this day \citep{chenenergyconserving2021M, aimechanical2021M}, and a clarification is needed.

In this work, we extend the results of ZK05, and clarify the misunderstandings in previous works on the MHD RS forming condition.
In Section~\ref{sec:rsproperty}, we present the theoretical basis that determines the properties of a magnetized RS.
In Section~\ref{sec:rscondition}, we discuss the different RS forming conditions in the literature.
We argue that the ZK05 condition, being consistent with the numerical simulations by MGA09, is the strict condition.
The GMA09 condition, though conceptually correct, is only an order-of-magnitude estimate of the actual condition given by ZK05.
Using the ZK05 condition, we define a RS forming radius, and discuss the dynamical evolution of the ejecta in the external shock stage in Section~\ref{sec:rsradius}.
Conclusions are given in Section~\ref{sec:conclusions}.

\section{Properties of the reverse shock}
\label{sec:rsproperty}

For a general astrophysical scenario where two flows of plasma collide with each other, the system can be approximately cast into a Riemann problem where nonlinear waves develop from the initial discontinuity of physical quantities.
A hydrodynamic Riemann problem admits a three-wave solution, i.e., a contact discontinuity in the middle and two nonlinear waves (shocks or rarefaction waves).
In this hydrodynamic limit, it is possible to determine the wave pattern (either two shocks, two rarefaction waves, or one shock and one rarefaction wave) from the initial condition prior to solving the problem \citep{2013rehy.book.....R}.
An MHD Riemann problem, on the other hand, generally admits a seven-wave solution including one contact discontinuity, two slow waves, two Alfven waves and two fast waves. Each of the fast or slow waves may either be a shock or a rarefaction wave.
This constitutes a much more complex problem for which there has been no practical method to 
predict the wave patterns before obtaining the complete numerical solution via iterations \citep{2006JFM...562..223G}.
However, a special case of MHD Riemann problem where the magnetic field is orthogonal to the flow velocity (i.e., the perpendicular configuration) could form a degeneracy so that slow waves and Alfven waves vanish \citep{1999MNRAS.303..343K}.

Here, we restrict ourselves to considering a perpendicular RS-FS system with an ambient medium of a constant number density (e.g., interstellar medium; ISM).
Following the convention of ZK05, we define four regions separated by three discontinuity surfaces: (1) FS upstream (unshocked), (2) FS downstream (shocked), (3) RS downstream (shocked), (4) RS upstream (unshocked).
Regions 2 and 3 are separated by a contact discontinuity.
$Q_{ij}$ represents the quantity $Q$ in region $i$ measured in the rest frame of $j$.

We define $p_\mathrm{tot,r}$ and $p_\mathrm{tot,f}$ as the total pressures (i.e., the sum of magnetic pressure and thermal pressure) in the shocked regions near the RS and FS, respectively.
Following the derivation in ZK05 (for a brief summary, see Appendix~\ref{sec:appendixa}), we find that the shock jump conditions yield
\begin{align}
    & p_\mathrm{tot,r} = (\hat{\Gamma}_3 \gamma_{43} +1)(\gamma_{43}-1)n_4 m_\mathrm{p} c^2 F_{43}\, ,\\
    & p_\mathrm{tot,f} = (\hat{\Gamma}_2 \gamma_{21} +1)(\gamma_{21}-1)n_1 m_\mathrm{p} c^2 F_{21}\, ,
    \label{eq:pfs}
\end{align}
where $\hat{\Gamma}$ is the adiabatic index given by the equation of state (EOS), $\gamma$ is the Lorentz factor associated with the dimensionless speed $\beta$, $n$ is the rest-frame number density, $m_\mathrm{p}$ is the proton mass, $c$ is the speed of light, $F_{21}=F_{21} (\gamma_{21},\sigma_1)$ and $F_{43} (\gamma_{43},\sigma_4)$ are the correction factors due to the magnetic field as detailed in Appendix~\ref{sec:appendixa} and ZK05.
For hydrodynamic shocks, we have $F_{21}=F_{43}=1$.
We assume the pressures near FS and RS balance each other, i.e., $p_\mathrm{tot,r} = p_\mathrm{tot,f}$, which is a reasonable approximation when the shocks initially form\footnote{
However, as the shocks propagate and develop, the two pressures differ, typically within an order of magnitude from each other \citep{kobayashiHydrodynamics1999A, beloborodovMechanical2006A, mimicaDeceleration2009A, uhmSemianalytic2011A, chenenergyconserving2021M, aimechanical2021M, aiEnginefed2022ae}.
}.
Assuming the pressure balance condition, we find strictly that (see also equation (32) of ZK05 for a special case)
\begin{equation}
    \frac{n_4 F_{43}}{n_1 F_{21}} = \frac{(\gamma_{21}-1)(\hat{\Gamma}_{2}\gamma_{21}+1)}{(\gamma_{43}-1)(\hat{\Gamma}_{3}\gamma_{43}+1)}\ .
    \label{eq:Fgeneral}
\end{equation}

Notice that to reach this conclusion, we have already assumed the unshocked regions (i.e., regions 1 and 4) are cold, such that the gas pressures in these regions are negligible.
We adopt a simple approximation of the adiabatic index
\begin{equation}
    \hat{\Gamma}_{2} = \frac{4\gamma_{21}+1}{3\gamma_{21}}\, ,\ \hat{\Gamma}_{3} = \frac{4\gamma_{43}+1}{3\gamma_{43}}\, ,
\end{equation}
which is suitable for an ideal gas in both non-relativistic and relativistic regimes \citep{uhmSemianalytic2011A}.
The FS region is nearly unmagnetized ($\sigma_2,\sigma_1 \ll 1$), and therefore we get $F_{21} \simeq 1$.
The relation~\eqref{eq:Fgeneral} then reduces to \citep[see also equation (4.147) of ][]{zhangPhysics2018TPoGBbBZI9CUP}
\begin{equation}
    F_{43}\frac{n_4}{n_1} = \frac{\gamma_{21}^2-1}{\gamma_{43}^2-1}\ .
    \label{eq:Ftransit}
\end{equation}

We treat the shocked region between RS and FS as a blastwave with uniform Lorentz factor ($\gamma_2=\gamma_3$), which is the basic assumption of mechanical models for both hydrodynamical problems \citep{beloborodovMechanical2006A} and MHD problems 
\citep{aimechanical2021M, aiEnginefed2022ae}.
This approximation is confirmed to be reasonable through both hydrodynamical simulations \citep{kobayashiHydrodynamics1999A} and MHD simulations \citep{mimicaDeceleration2009A, aiEnginefed2022ae}.
According to the Lorentz transformation $\gamma_{21} = \gamma_{43}\gamma_4(1-\beta_{43}\beta_4)$ and $\gamma^2-1=\gamma^2\beta^2$, relation~\eqref{eq:Ftransit} is equivalent to
\begin{equation}
    \gamma_{41}^2 \frac{n_1}{n_4} = \left[\frac{\beta_{43}}{\beta_{21}(1-\beta_{43}\beta_4)}\right]^2 F_{43}\, .
\end{equation}
We assume a relativistic ejecta and an FS that is at least mildly relativistic with $\beta_4,\beta_{21} \sim 1$ (typical for the GRB case), and we find
\begin{equation}
    \gamma_{4}^2 \frac{n_1}{n_4} = \left(\frac{\beta_{43}}{1-\beta_{43}}\right)^2 F_{43}\, .
    \label{eq:zetarelation}
\end{equation}
Define a parameter
\begin{equation}
    \zeta \equiv \gamma_4^2 \frac{n_1}{n_4}\, ,
    \label{eq:zetadef}
\end{equation}
which is only determined by the initial conditions of the unshocked regions.
Noticing 
that $F_{43}$ is a function of $\sigma$ and $\beta_{43}$ only (see Appendix~\ref{sec:appendixa})
, we find from equation (\ref{eq:zetarelation}) that $\zeta$ can also be expressed as $\zeta = \zeta (\sigma, \beta_{43})$.
Therefore, when the initial conditions $(\zeta, \sigma)$ of the unshocked regions are known, the relative Lorentz factor 
between the upstream and the downstream of the RS, i.e., $\gamma_{43}$, 
is determined straightforwardly from the relation (\ref{eq:zetarelation}).
Hereafter, we define 
\begin{equation}
    \gamma_\mathrm{rel}^\mathrm{RS} \equiv \gamma_{43}\, .
    \label{eq:garel}
\end{equation}

Motivated by observations and the discussion in MGA09, we investigate the dissipative efficiency of the RS.
We assume that all the internal energy behind the shock is carried by the accelerated particles and converted into radiation, as discussed in \citet{narayanConstraints2011M}, \citet{komissarovShock2012M} and \citet{sironiRelativistic2015M}.
This is highly optimized so that the efficiency values obtained here are the maximum values.
The dissipative efficiency of the RS is defined as the fraction of the total enthalpy that is converted into internal energy \citep[for slightly different definitions, see][]{narayanConstraints2011M, komissarovShock2012M, sironiRelativistic2015M}, i.e.,
\begin{equation}
\begin{split}
    f_\mathrm{sh} & \equiv \frac{e_3}{(n_3 m_\mathrm{p} c^2 + e_3 + p_3)(1+\sigma_3)}\\
    & = \frac{(\gamma_{43}-1)f_{a,43}}{\left[1 + \hat{\Gamma}_{3}(\gamma_{43}-1)f_{a,43}\right](1+\sigma_3)}\, ,
    \label{eq:fsh}
\end{split}
\end{equation}
which is a function of $\gamma_{43}$ and $\sigma$.
Here, $f_a$ is a correction factor as detailed in Appendix~\ref{sec:appendixa} and ZK05.
As discussed in the last paragraph, the relative Lorentz factor $\gamma_{43}$ is determined by the conditions $(\zeta, \sigma)$ of the unshocked regions.
Therefore, the dissipative efficiency can be calculated as long as $(\zeta, \sigma)$ are known.

\section{Reverse shock forming condition: Reconciling theories and simulations}
\label{sec:rscondition}

\begin{figure}
	\includegraphics[width=\columnwidth]{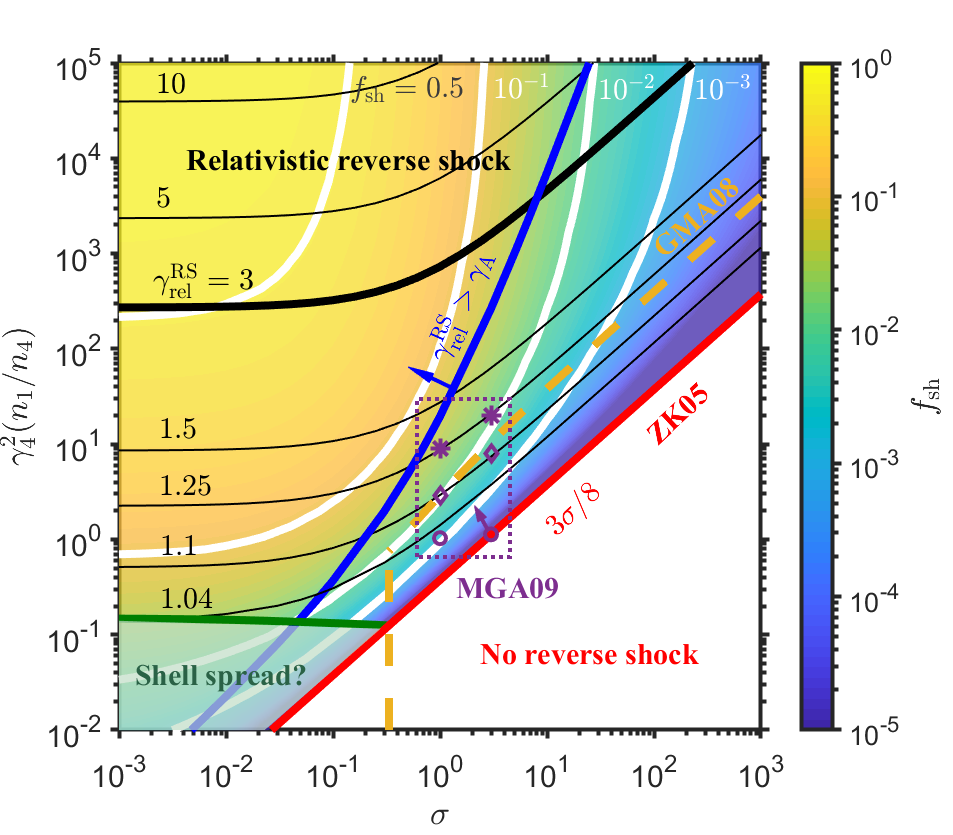}
    \caption{
    RS properties in the parameter space of $\zeta = \gamma_4^2 (n_1/n_4)$ and magnetization $\sigma$.
    We show the contours of the RS relative Lorentz factor $\gamma_\mathrm{ rel}^\mathrm{ RS}$ (black solid lines labeled on the left side), and the contours of the RS dissipative efficiency $f_\mathrm{sh}$ (colored region with values indicated by the colorbar; $f_\mathrm{ sh} = 0.5,10^{-1},10^{-2},10^{-3}$ are highlighted in white solid lines labeled on the top).
    The quantities $\gamma_\mathrm{ rel}^\mathrm{ RS}$ and $f_\mathrm{sh}$ are defined in equations~\eqref{eq:garel} and \eqref{eq:fsh}, respectively.
    The RS forming conditions given by ZK05 (see Sec.~\ref{sec:ZK05}), \citet{lyutikovSimple2010PRE, lyutikovDynamics2011M} (see Sec.~\ref{sec:Lyutikov}), and GMA08 (see Sec.~\ref{sec:GMA08}) are indicated by the red solid line, the blue solid line, and the orange dashed line, respectively.
    Six purple markers in the dotted frame represent six data points taken from the numerical simulations of MGA09 (see Sec.~\ref{sec:MGA09}).
    Specifically, the circles, diamonds, and asterisks represent the parameters $(\zeta, \sigma)$ when the relative Lorentz factor $\gamma_\mathrm{ rel}^\mathrm{ RS}$ becomes $1.04, 1.1, 1.25$ according to MGA09, respectively.
    The arrow on the last circle indicates that its actual location on the diagram should be moved upwards to the left due to shell expansion, making it more consistent with our prediction (as discussed in Sec.~\ref{sec:MGA09}).
    Below the dark green solid line at the bottom left, the shell spreading can further complicate the problem (see Sec.~\ref{sec:GMA08} and \ref{sec:rsradius} for discussions).
    }
    \label{fig:1}
\end{figure}

In the parameter space of unshocked parameters $(\zeta, \sigma)$ in Fig.~\ref{fig:1}, we show the contours of the relative Lorentz factor $\gamma_\mathrm{ rel}^\mathrm{ RS}$ of the RS and the contours of RS efficiency $f_\mathrm{ sh}$.
They are calculated using equation~\eqref{eq:zetarelation} and equation~\eqref{eq:fsh}, respectively.
Black solid lines represent the lines of a constant relative Lorentz factor $\gamma_\mathrm{ rel}^\mathrm{ RS}$, labeled by their values on the left side of the diagram.
The $\gamma_\mathrm{ rel}^\mathrm{ RS} = 3$ curve is highlighted in bold, as the approximate boundary between near-Newtonian RS and RRS.
The dissipative efficiencies $f_\mathrm{ sh}$ are indicated by colors shown in the colorbar, with $f_\mathrm{ sh} = 0.5,10^{-1},10^{-2},10^{-3}$ highlighted in white solid lines labeled by their values on the top.

For a weakly-magnetized ejecta ($\sigma \ll 1$), the strength of the RS is solely determined by the parameter $\zeta = \gamma_4^2 (n_1/n_4)$.
The RS becomes relativistic at $\zeta \gg 1$ and Newtonian at $\zeta \ll 1$, as proposed in the early work of \citet{sariHydrodynamic1995A}.
The RRS is efficient at dissipating energy, whilst the Newtonian RS is not.
The highest efficiency is reached in the case of a non-magnetized RRS, where we have $f_\mathrm{sh} \simeq 3/4$.

The increase in the magnetization degree $\sigma$ effectively weakens the RS, as the relative Lorentz factor of the RS can be approximately solved using the relation $\zeta \sim \gamma_{43}^4 \beta_{43}^2 (1+\sigma)$.
The dissipative efficiency is further suppressed in the high-$\sigma$ regime because a large portion of magnetic energy cannot be converted into internal energy.
For any RS with $\sigma \gtrsim 1$, it dissipates energy at a low efficiency ($f_\mathrm{sh} \lesssim 0.01-0.1$).
As widely recognized, this suggests that the RS emission may become too weak to be observed for $\sigma \gtrsim 1$ \citep{zhangGamma2005Aa, gianniosexistence2008A, mimicaDeceleration2009A, granotInteraction2012M, zhangPhysics2018TPoGBbBZI9CUP}.

To reconcile theories and simulations on RS formation, we integrate the results from different works in Fig.~\ref{fig:1}, as discussed below separately.

\subsection{Treatment of ZK05}
\label{sec:ZK05}

A shock exists only if the upstream total pressure exceeds the downstream total pressure \citep{1967rhm..book.....L, 1987PhFl...30.3045M, 2013rehy.book.....R}.
At the critical point where pressure balance is established between the upstream and downstream, the shock vanishes and the two regions merge into one continuous flow.
When the RS transits to a rarefraction wave, the total pressure at the RS is simply the the magnetic pressure in the unshocked region 4, i.e., $p_\mathrm{tot,r} \rightarrow 2\sigma n_4 m_\mathrm{p} c^2$.
The total pressure at the FS is given by the jump condition~\eqref{eq:pfs} with $\gamma_{21}$ approaching $\gamma_4$, i.e., $p_\mathrm{tot,f} \rightarrow (4/3)(\gamma_4 \beta_4)^2 n_1 m_\mathrm{p} c^2$.
The RS forming condition for a relativistic ejecta is thus 
\begin{equation}
    \gamma_4^2 \frac{n_1}{n_4} > \frac{3}{8} \sigma\, .
    \label{eq:ZK05}
\end{equation}
In the 1D planar symmetry, this condition is supported by the numerical simulations \citep{2009ApJ...690L..47M}.
This criterion given in ZK05 is shown in Fig.~\ref{fig:1} as a red solid line, below which there is no RS.

\subsection{Treatment of Lyutikov}
\label{sec:Lyutikov}

\citet{lyutikovSimple2010PRE, lyutikovDynamics2011M} argued that although the criterion~\eqref{eq:ZK05} holds in 1D problems, the RS may not form in multi-dimensions if the CD is subsonic with respect to the ejecta. He found that this condition can be expressed asymptotically as
\begin{equation}
    \gamma_4^2 \frac{n_1}{n_4} > 
    \begin{cases}
    3\sigma/8\, , \sigma \ll 1\\
    6 \sigma^3\, , \sigma \gg 1
    \end{cases}
    .
    \label{eq:lyutikov}
\end{equation}
A rigorous description of this condition is to have $\gamma_\mathrm{rel}^\mathrm{RS} > \gamma_\mathrm{A}$, where $\gamma_\mathrm{A} = \sqrt{1+\sigma}$ is the Lorentz factor of the Alfven speed (the speed of the fast magnetosonic wave for cold flows).
In Fig.~\ref{fig:1}, we plot the critical condition where $\gamma_\mathrm{rel}^\mathrm{RS} = \gamma_\mathrm{A}$ as the blue solid curve, which approaches the asymptotic expression~\eqref{eq:lyutikov} at both ends.

There has been no evidence to support this tighter constraint from multi-dimensional simulations on relativistic MHD (RMHD) shocks \citep[e.g., ][]{1999MNRAS.303..343K, 2005JFM...544..323R, 2006JFM...562..223G}, and the GRB afterglow-related simulations were mostly restricted to 1D \citep[e.g., ][]{mimicaDeceleration2009A, 2009ApJ...690L..47M}.
Nevertheless, in Fig.~\ref{fig:1} we confirm the statement of \citet{lyutikovSimple2010PRE, lyutikovDynamics2011M} that the RS, even if it exists when $\gamma_\mathrm{rel}^\mathrm{RS} < \gamma_\mathrm{A}$, is `weak', in a sense that it can not dissipate energy efficiently as a near-Newtonian or strongly-magnetized shock.

\subsection{Treatment of GMA08}
\label{sec:GMA08}

GMA08 considered a magnetized shell with the width $\Delta$ and the bulk Lorentz factor $\gamma_4$ at radius $r$, all measured in the lab frame\footnote{
The shell-spreading effect in spherical geometry is not taken into account in this section.
Therefore, we use the notation $\Delta$ and the initial shell width $\Delta_0$ interchangeably in this section, unless otherwise noted.
}.
The strength of the RS was conventionally parameterized by
\begin{equation}
    \xi \equiv \sqrt{l/\Delta}\gamma_4^{-4/3}\, ,
    \label{eq:xi}
\end{equation}
where $l \equiv (3E/4\pi n_1 m_\mathrm{ p}c^2)^{1/3}$ is the Sedov length associated with the total energy of the ejecta $E=4\pi r^2 \Delta \gamma_4^2 n_4 m_\mathrm{ p} c^2 (1+\sigma)$, and $n_4$ is the rest-frame particle number density of the shell at $r$.

GMA08 claimed that the RS only forms when the RS crosses the shell before the casual contact is established through MHD waves, i.e., $r_\mathrm{ c} > r_\Delta$.
Here, $r_\mathrm{ c}$ is the `contact' radius where MHD waves cross the shell on the expansion timescale $r/\gamma_4 c$, and $r_\Delta$ is the RS crossing radius where the RS crosses the shell.
The radii are given by (see equation (8) of GMA08 and equation (38) of ZK05)
\begin{align}
    & r_\mathrm{ c} \sim \Delta \gamma_4^2 \left(\sqrt{\frac{1+\sigma}{\sigma}}-1\right)\, , 
    \label{eq:rc}\\
    & r_\Delta \sim  \Delta \gamma_4^2 \zeta^{-1/2} (1+\sigma)^{-1/2}\, .
    \label{eq:rrs}
\end{align}
Consequently, the RS forming condition in GMA08 can be cast into
\begin{equation}
    \gamma_4^2 \frac{n_1}{n_4} \gtrsim  (1+\sigma)^{-1}\left(\sqrt{\frac{1+\sigma}{\sigma}}-1\right)^{-2}
    \label{eq:GMA08}
\end{equation}
when the shell does not spread.
Note that here the quantity $\zeta$ is evaluated at the RS crossing radius $r_\Delta$.
If the shell-spreading is taken into account, GMA08 argued that the RS always exists as long as the contact radius is larger than the shell-spreading radius $r_\mathrm{ s} \sim \Delta_0 \gamma_4^2$, which is equivalent to $\sigma \lesssim 1/3$.
This criterion is plotted in Fig.~\ref{fig:1} as an orange dashed line, which is given by equation~\eqref{eq:GMA08} in the $\sigma \gtrsim 1/3$ regime and goes straight down in the $\sigma \lesssim 1/3$ regime due to the shell-spreading effect.
GMA08 claimed that there is no RS on the right hand side of this line in the parameter space.

Here, we argue that the ZK05 condition~\eqref{eq:ZK05} and the GMA08 condition~\eqref{eq:GMA08} are of the same physical origin as noticed in \citet{2009ApJ...690L..47M}, with the former being the strict one and the latter being an order-of-magnitude estimate.
This conclusion is reached based on two observations:
\begin{itemize}
    \item[$\bullet$]
    As the RS grows weak and vanishes, the upstream total pressure balances the downstream total pressure, \textit{and at the same time} the shock speed decreases and becomes equal to the magnetosonic wave speed (see, e.g., Section 4.4.3 in \citet{2013rehy.book.....R} for relativistic hydrodynamic shocks, Figure 20.5 in \citet{goedbloedAdvanced2010AM} for Newtonian MHD shocks, Figure 2 in \citet{2009ApJ...690L..47M} for RMHD shocks).
    The former is the theoretical basis for ZK05 condition, and the latter is the starting point for GMA08.
    Therefore, these two physical concepts can both serve as the RS forming condition\footnote{
    \citet{granotInteraction2012M} criticized the GMA08 condition for violating the fact that fast shocks travel faster than linear waves.
    Here, we point out that although his observation is correct, it is important to acknowledge that the fast shock reduces to a linear wave when the shock is at the critical point to vanish.
    }.
    \item[$\bullet$]
    Both criteria show the same scaling with $\sigma$ in Fig.~\ref{fig:1}.
    The GMA08 condition~\eqref{eq:GMA08} gives $\zeta \gtrsim 4\sigma$ when $\sigma \gg 1$, which is similar to the ZK05 condition~\eqref{eq:ZK05} but roughly an order of magnitude higher\footnote{
    This is supported by the same scaling law of critical $\sigma$ with the bulk Lorentz factor, ejecta width, and total energy in ZK05 and GMA08.
    Comparing the scaling law (43) in ZK05 and (11) in GMA08, we find that they express the same scaling relation, but the numerical factors are different.
    }.
    We attribute this discrepancy to the approximation used in GMA08 to estimate the characteristic radii, which omits the detailed numerical factors.
    As a proof, we provide a rigorous derivation of the contact radius $r_\mathrm{ c}$ and the RS crossing radius $r_\Delta$ in Appendix~\ref{sec:appendixb}, and explain that they can only become equal to each other when the pressure balance is established at the ZK05 condition~\eqref{eq:ZK05}.
\end{itemize}

As a comparison to equation~\eqref{eq:GMA08}, GMA08 wrote the condition in terms of $\xi$ as $\xi^3 \lesssim (1+\sigma)(\sqrt{(1+\sigma)/\sigma}-1)^2$.
The dimensionless parameters can thus be converted into each other through a simple relation at the RS crossing radius $r_\Delta$ as
\begin{equation}
    \zeta \sim \xi^{-3}\, .
    \label{eq:xitozeta}
\end{equation}
It should be pointed out here that all the derivations and simulations of GMA08 and MGA09 (see Sec.~\ref{sec:MGA09}) are based on the assumption that the $\xi$ parameter (converted to $\zeta$ in our context) is evaluated at the RS crossing radius $r_\Delta$.
Only under this assumption can the relation $\zeta \sim \xi^{-3}$ hold.
As discussed in more detail in Sec.~\ref{sec:sigcr}, we argue that the parameters evaluated at $r_\Delta$ should not be used to track the evolution of the shell or to determine the RS forming condition\footnote{
For reference, ZK05 gave the critical $\sigma_\mathrm{cr}$ based on the parameters evaluated at the deceleration radius, which we find is also not a satisfying answer.
}.
To track the evolution of the shell, we should use the parameters $\xi$ (or $\zeta$) evaluated at the radius comoving with the shell.
To estimate the critical $\sigma_\mathrm{cr}$ at which the RS is formed, we should use the parameter $\xi$ (or $\zeta$) evaluated at the radius where the critical RS forming condition is satisfied.

Here, we assume that the shell does not spread.
If the shell spreading is taken into account, we need to distinguish between the initial parameters $(\zeta_0, \sigma_0)$ before shell spreading and the parameters $(\zeta, \sigma)$ after shell spreading\footnote{
It is important to highlight at this point that the diagram in Fig.~\ref{fig:1} only holds when the parameters $(\zeta, \sigma)$ are taken in the unshocked regions \textit{right behind} the RS (or the reverse rarefaction wave when the RS does not form).
}.
GMA08 suggested that when $\sigma_0 \lesssim 1/3$, the RS always forms due to the shell-spreading effect.
Although we find that the RS will always form eventually for $\sigma_0 \lesssim 1$ if shell spreading is taken into account (see Sec.~\ref{sec:rsradius}), we do not find the breaking of causal contact at $\sigma_0 \lesssim 1/3$ (as suggested by GMA08) a convincing argument for the formation of RS.
This is discussed more thoroughly in Sec.~\ref{sec:compare}.

Motivated by GMA08, we further explore the regime where the shell could spread before the RS crosses the shell, i.e., $r_\mathrm{s} < r_\Delta$.
We adopt a more rigorous expression~\eqref{eq:rrsrigorous} of $r_\Delta$, and compute the ratio of the RS crossing radius $r_\Delta$ to the shell-spreading radius $r_\mathrm{s}$.
In Fig.~\ref{fig:1}, we plot the critical condition where $r_\mathrm{s} = r_\Delta$ as the dark green solid line.
We find that below this line, the shell spreads before the RS crosses the shell ($r_\mathrm{s} < r_\Delta$; `thin shell' regime discussed in Sec.~\ref{sec:thinshell}).
All the weakly-magnetized thin shells that have undergone shell spreading are expected to enter this regime and form RSs which cross the shell as trans-relativistic shocks with $\gamma_\mathrm{rel}^\mathrm{RS}\simeq 1.04$ (see a detailed discussion in Sec.~\ref{sec:rsradius}).
Nevertheless, for a wide range of parameter space above the dark green solid line and the red solid line, the RS crosses the shell before shell spreading ($r_\mathrm{s} > r_\Delta$; `thick shell' regime discussed in Sec.~\ref{sec:thickshell}).
We refer to Sec.~\ref{sec:rsradius} for a more detailed discussion on shell spreading.


\subsection{Treatment of MGA09}
\label{sec:MGA09}

MGA09 performed high-resolution simulations in 1D spherical geometry, focusing on the evolution of RS and FS in RMHD.
They confirmed the finding of GMA08 that the evolution of the system is characterized by two parameters, namely $\xi$ and $\sigma$.
Using the relation~\eqref{eq:xitozeta} to convert $\xi$ into $\zeta$, we plot six simulated data points taken from Figure 4 of MGA09, shown as six purple markers highlighted by a dotted frame in Fig.~\ref{fig:1}.
The models in MGA09 are taken for $\sigma = 1, 3$ when the relative Lorentz factor $\gamma_\mathrm{rel}^\mathrm{RS}$ becomes $1.04$ (circles), $1.1$ (diamonds), and $1.25$ (asterisks), respectively.

Our predictions of the relative Lorentz factors fit perfectly with the simulated results of MGA09 for $\gamma_\mathrm{rel}^\mathrm{RS} = 1.1, 1.25$.
Therefore, we conclude that the theoretical formulation of ZK05 is consistent with the simulations of MGA09 if the RS is not too weak ($\gamma_\mathrm{rel}^\mathrm{RS} \gtrsim 1.1$).
We confirm the results of MGA09 that the magnetized weak RS (circles and diamonds) dissipates less than $\sim 1\%$ of the total energy, indicating a suppressed emission.

However, our theory underestimates the relative Lorentz factor for the $\zeta\sim 1.1,\sigma=3,\gamma_\mathrm{rel}^\mathrm{RS}=1.04$ model, which we predict to be at the verge of vanishing.
The discrepancy for near-Newtonian RS ($\gamma_\mathrm{rel}^\mathrm{RS} \lesssim 1.1$) may be attributed to the internal evolution of the ejecta profile in spherical geometry, which is not captured by our theory.
For instance, Figure 1 in MGA09 demonstrated that the rarefaction wave propagating backwards results in the radial expansion of the shell.
This subsequently leads to decreases in both the number density $n_4$ and the magnetization degree $\sigma$, creating an otherwise impossible near-Newtonian RS in the rarefaction fan.
Therefore, it becomes important in these cases to define the initial parameters $\zeta_0$ and $\sigma_0$ to be distinguished from $\zeta$ and $\sigma$ right in front of the RS.
We expect that the actual location of the near-Newtonian RS (circles) on Fig.~\ref{fig:1} should be moved upwards to the left (indicated by the purple arrow), where $\sigma < \sigma_0$ and $\zeta > \zeta_0$ due to the radial expansion of the shell.
This would explain the underpredicted relative Lorentz factor of the near-Newtonian RS.
For stronger RS ($\gamma_\mathrm{rel}^\mathrm{RS} \gtrsim 1.1$), the RS has either propagated to the end of the rarefaction fan or formed at the CD, which limits the effect of shell expansion induced by rarefaction waves.

\section{Reverse shock forming radius: an order-of-magnitude view of shell evolution}
\label{sec:rsradius}

\begin{figure}
	\includegraphics[width=\columnwidth]{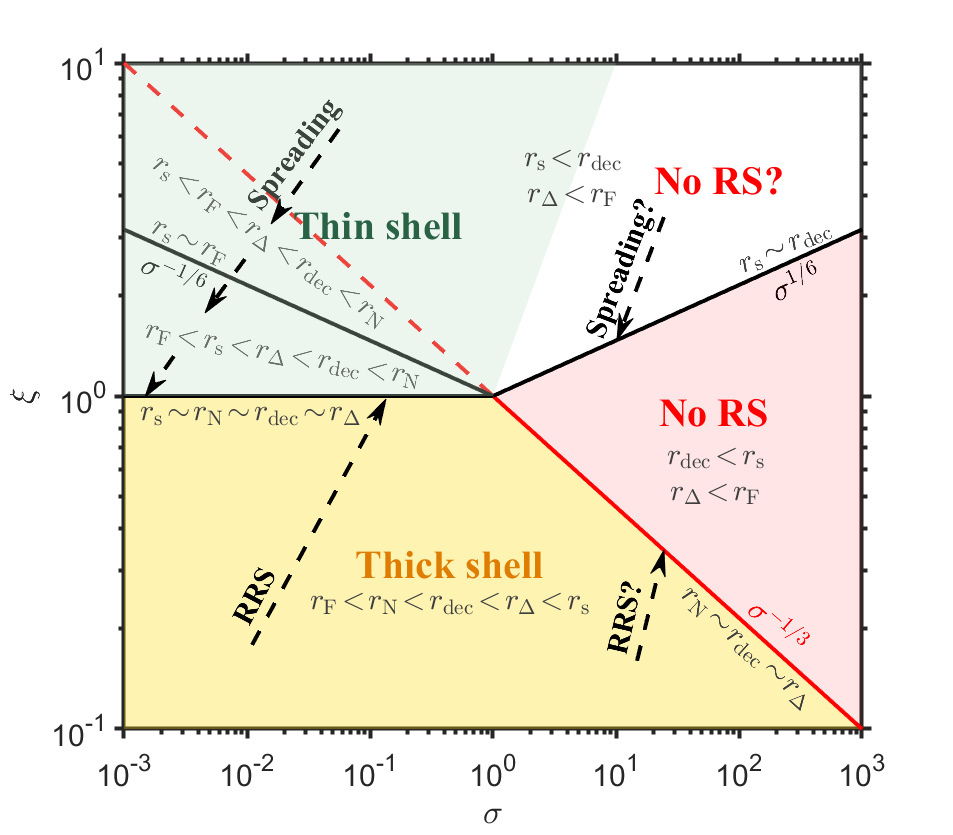} 
    \caption{
    Order-of-magnitude classification for the dynamical evolution of the shell in the parameter space of $\xi = \sqrt{l/\Delta}\gamma_4^{-4/3}$ and magnetization $\sigma$.
    Yellow region: Generalized thick shell regime.
    Green region: Generalized thin shell regime.
    Red region: Direct deceleration regime with no RS.
    White region: Uncertain regime where the shell spreads and decelerates with no RS.
    The approximate critical condition $\xi \sim \sigma^{-1/3}$ for the RS formation is highlighted in red, which separates the RS forming regime (left to the red line; $r_\mathrm{F}<r_\Delta$) and the initially no RS regime (right to the red line; $r_\Delta<r_\mathrm{F}$), consistent with Figure 1 in GMA08.
    In the red region, the shell is directly decelerated by the FS with no RS or shell spreading.
    In the yellow region, the RS is formed, becomes an RRS, and fully decelerates the shell when it reaches the rear part of the shell at the double coincidence of $r_\Delta \sim r_\mathrm{dec}$, which is the conventional `thick shell' scenario in hydrodynamics \citep{sariHydrodynamic1995A}.
    In the green region, shell spreading causes decreases in both $\xi$ and $\sigma$, leading to the formation of the RS which strengthens into a trans-relativistic shock at the triple coincidence $r_\mathrm{N} \sim r_\Delta \sim r_\mathrm{dec}$ \citep[the `thin shell' scenario in hydrodynamics;][]{sariHydrodynamic1995A}.
    The evolution of shell parameters is indicated by the dashed arrows, attributed to the shell spreading in the `thin shell' regime and the deceleration by RRS in the `thick shell' regime, respectively.
    In the green region on the right of the red dashed line, the RS will not form if the subsequent shell spreading is omitted.
    However, in the high-$\sigma$ thin shell regime (white region), how the shell responds to the shell-spreading effect may determine whether it forms a RS or directly enters the deceleration phase, which is not well understood.
    We expect that the shell may spread and then be decelerated by the FS without forming a RS in the white region.
    }
    \label{fig:2}
\end{figure}

Using the pressure balance condition for RS formation, we can define a new characteristic radius, i.e., the RS forming radius $r_\mathrm{F}$ where the RS is formed.
This can be done by substituting the number density $n_4$ with the expression for total energy $E=4\pi r^2 \Delta \gamma_4^2 n_4 m_\mathrm{ p} c^2 (1+\sigma)$ in ZK05 condition~\eqref{eq:ZK05}, which gives \citep[see also equation (7) in][although in a different context]{mimicaDeceleration2009A}
\begin{equation}
    r_\mathrm{F} \equiv \left(\frac{3E\sigma}{32\pi n_1 m_\mathrm{p} c^2 \gamma_4^4 \Delta (1+\sigma)}\right)^{1/2}\, .
\end{equation}

Following ZK05, we can now estimate five characteristic radii relevant for the external shocks:
(1) the shell-spreading radius\footnote{
Although the shell spreading in high-$\sigma$ regime has not been carefully explored, there is evidence that the shell-spreading radius can still be estimated as $r_\mathrm{s} \sim \gamma_0^2 \Delta_0$ in highly-magnetized ejecta.
The early analytical treatment of the hydrodynamics of ejecta suggested that the shell spreading arises from the spherical geometry during the coasting phase of the cold ejecta \citep{meszarosGasdynamics1993A, piranHydrodynamics1993M}.
If the coasting stage still exists for highly-magnetized ejecta, we expect that the expression for the shell-spreading radius would be the same.
However, if there is no coasting phase in high-$\sigma$ regime \citep[see, e.g.,][]{granotImpulsive2011M, granotInteraction2012M, gaoPhotosphere2015A}, it is unclear if the shell can spread at all.
} $r_\mathrm{s} \sim \gamma_0^2 \Delta_0$, (2) the RS forming radius $r_\mathrm{F} \sim l^{3/2} \Delta^{-1/2} \gamma_4^{-2} \sigma^{1/2} (1+\sigma)^{-1/2}$, (3) $r_\mathrm{N} \sim l^{3/2} \Delta^{-1/2} \gamma_4^{-2}$ where the RS transforms from a Newtonian shock to a relativistic shock, (4) the RS crossing radius $r_\Delta \sim \Delta^{1/4} l^{3/4} (1+\sigma)^{-1/2}$ where the RS reaches the rear part of the shell, and (5) the deceleration radius $r_\mathrm{dec} \sim l \gamma_4^{-2/3} (1+\sigma)^{-1/3}$ where the FS collects $1/\gamma_4$ of the shell rest mass\footnote{
GMA08 defined the deceleration radius as $r_\mathrm{ dec} = l/\gamma_4^{2/3}$.
However, ZK05 argued that it should be defined using the kinetic energy alone, as the differences in magnetic energy is negligible.
For a discussion of this point, see \citet{lyutikovComment2005ae} and \citet{zhangReply2005ae}.
At ZK05 deceleration radius, a magnetized shell may decelerate with a shallow deceleration index $\gamma_4 \propto r^{-1/2}$ \citep[for ISM;][]{zhangPhysical2006A, granotImpulsive2011M, granotInteraction2012M}, and later on transforms into a Blandford-Mckee self-similar deceleration phase at the deceleration radius defined by the total energy \citep{blandfordFluid1976PF, zhangPhysics2018TPoGBbBZI9CUP}.
Here, we follow the definition of ZK05.
We find these two different definitions lead to similar classifications of the shell evolution shown in Fig.~\ref{fig:2}.
}.
They are related to each other by (see also equation (51) in ZK05 for the first four radii)
\begin{equation}
    \xi_0^2 r_\mathrm{s} 
    = (1+\sigma)^{1/3} r_\mathrm{dec} 
    = \xi^{1/2} (1+\sigma)^{1/2} r_\Delta 
    = \frac{r_\mathrm{N}}{\xi}
    = \left(\frac{1+\sigma}{\sigma}\right)^{1/2} \frac{r_\mathrm{F}}{\xi}\, ,
    \label{eq:radiirelation}
\end{equation}
where $\xi_0$ is the parameter $\xi$ defined by the initial width $\Delta_0$ and the initial Lorentz factor $\gamma_0$.
Notice that only the shell-spreading radius $r_\mathrm{s}$ is defined by the initial conditions, whereas the other four radii can in principle change with the shell parameters as the shell evolves.
In the parameter space of $(\xi, \sigma)$, there are multiple regions representing different physical processes.
Two features can be used to simplify the problem:
(1) $r_\Delta \lesssim r_\mathrm{F}$ ($\xi \gtrsim \sigma^{-1/3}$; right hand side of the red line in Fig.~\ref{fig:2}) gives the unphysical scenario where the RS crosses the shell even before the RS can be formed, which in fact represents the regime where RS cannot be formed initially.
In these cases, the characteristic radii $r_\mathrm{F}, r_\mathrm{N}, r_\Delta$ associated with the RS lose their physical meanings, and thus should be left out of our discussion.
(2) $r_\mathrm{F}$ is always smaller than $r_\mathrm{N}$ but approaches $r_\mathrm{N}$ in the high-$\sigma$ regime.


For the following discussion, we only focus on the external shocks.
We start with a shell propagating through the ISM at radius $r_0$ from the central engine, with all the five characteristic radii well beyond $r_0$. 
We assume that the shell is in a quasi-coasting stage, where the Lorentz factor $\gamma_0$, shell width $\Delta_0$, and magnetization $\sigma_0$ remain approximately constant as the shell moves\footnote{
This assumption is valid for weakly-magnetized shells.
For highly-magnetized shells, however, the ejecta may never reach a steady coasting phase.
The shell could directly enter the deceleration phase after the rapid acceleration if the coasting radius is larger than the deceleration radius \citep[see, e.g.,][]{granotImpulsive2011M, granotInteraction2012M, gaoPhotosphere2015A}.
Here, we limit our discussion to the cases where the coasting radius is smaller than all the five characteristic radii.
}.
Under this assumption, the parameter space in Fig.~\ref{fig:2} is generally separated into three regimes: the `thick shell' regime (Sec.~\ref{sec:thickshell}), the `thin shell' regime (Sec.~\ref{sec:thinshell}), and the `no RS' regime (Sec.~\ref{sec:nors}), as discussed below.

\subsection{`Thick shell' regime (yellow region)}
\label{sec:thickshell}

The RS is formed and becomes relativistic, which then fully decelerates the shell when it crosses the entire ejecta.

This regime is characterized by $r_\mathrm{F}<r_\mathrm{N}<r_\mathrm{dec}<r_\Delta<r_\mathrm{s}$.
The shell coasts freely as the number density drops with radius as $n_4 \propto r^{-2}$ until the RS forming condition~\eqref{eq:ZK05} is met.
A Newtonian RS is first formed at $r\sim r_\mathrm{F}$, which does not influence the dynamics of the shell appreciably ($\xi \simeq \xi_0,\sigma \simeq \sigma_0$).
As the shell propagates further, the number density continues to drop until the RS becomes relativistic at $r\sim r_\mathrm{N}$.
At $r\gtrsim r_\mathrm{N}$, the shell begins to be decelerated by the RRS, leading to a decrease in the Lorentz factor estimated to be $\gamma_4 \simeq l^{3/4} \Delta^{-1/4} r^{-1/2}$ \citep{sariVariability1997A, kobayashiHydrodynamics1999A}, although it is less clear whether this scaling holds in the high-$\sigma$ regime.
With this expression, we find that $r_\mathrm{N}$ in relation~\eqref{eq:radiirelation} can be replaced by the distance $r$, which gives
\begin{equation}
    \frac{r}{\xi^{3/2}}
    =\frac{(1+\sigma)^{1/3}}{\xi^{1/2}}r_\mathrm{dec}
    =(1+\sigma)^{1/2}r_\Delta
    = l^{3/4}\Delta_0^{1/4}, \ r_\mathrm{N} \lesssim r \lesssim r_\Delta\, .
\end{equation}
Therefore, when the shell passes $r_\mathrm{N}$, $\xi$ begins to increase as $\xi \propto r^{2/3}$, accompanied by the increase in $\sigma$ ($\sigma \propto r, \sigma \ll 1$ or $\sigma \propto r^{1/2}, \sigma \gg 1$; see Table~\ref{tab:1})\footnote{
Here, we assume the ejecta is decelerated by the RRS, leading to $\gamma_4 \propto r^{-1/2}$ even in the high-$\sigma$ regime.
Under this assumption, both the magnetic energy density and the rest mass density decrease with $r$.
However, the rest mass density has a steeper dependence on $r$, thus decreasing faster than the magnetic energy density.
This results in the increase in $\sigma$.
}.
This evolution is indicated by the arrows pointing upwards to the right in Fig.~\ref{fig:2}.
Following the arrow, $\xi$ continues to increase as the shell propagates to larger distances.
However, as long as the parameters stay in the yellow region, we always have $r<r_\mathrm{dec}<r_\Delta$.
This means that the shell has to continue travelling until reaching the boundary $\xi \sim (1+\sigma)^{-1/3}$ of the yellow region, where $r \sim r_\mathrm{dec} \sim r_\Delta$.

In summary, in this `thick shell' regime, a shell that begins with $\xi_0 < (1+\sigma_0)^{-1/3}$ forms an RS which strengthens into an RRS. Subsequently, the parameters $(\xi,\sigma)$ start to travel within the regime as the RRS decelerates the shell, until the shell adjusts itself to $\xi \sim (1+\sigma)^{-1/3}$ so that the RRS crosses and fully decelerates the shell at the same time (at the double coincidence of $r_\mathrm{dec}\sim r_\Delta$).

\subsection{`Thin shell' regime (green region)}
\label{sec:thinshell}

A Newtonian RS is formed either from the beginning or during shell spreading.
The shell is decelerated by both the RS and FS as the RS becomes trans-relativistic and crosses the shell at approximately the same time.
This regime can be further divided into two sub-regimes according to whether the shell spreads before the RS formation, as discussed below.

Above the black line $\xi \sim \sigma^{-1/6}$, $r_\mathrm{s}$ is the smallest radius.
The coasting phase persists until the shell begins to spread at $r\sim r_\mathrm{s}$ with $\Delta \sim r/\gamma_0^2$.
As long as the shell spreading is initiated, $r_\mathrm{s}$ in relation~\eqref{eq:radiirelation} can be replaced by the distance $r$, i.e.,
\begin{equation}
    \xi^2 r
\!=\! (1+\sigma)^{1/3} r_\mathrm{dec} 
\!=\! \xi^{1/2} (1+\sigma)^{1/2}\! r_\Delta 
\!=\! \frac{r_\mathrm{N}}{\xi}
\!=\! \left(\frac{1\!+\!\sigma}{\sigma}\right)^{1/2}\! \frac{r_\mathrm{F}}{\xi}
\!=\! l\gamma_0^{-2/3}\, .
\label{eq:radiirelationspread}
\end{equation}
At $r \gtrsim r_\mathrm{s}$, both $\xi$ and $\sigma$ decrease as the shell propagates outwards (indicated by the arrow pointing downwards to the left in Fig.~\ref{fig:2}; see Table~\ref{tab:1}).
A Newtonian RS is formed at $\xi\sim \sigma^{-1/6}$, where $r\sim r_\mathrm{F}$.
Nevertheless, we always find $r<r_\Delta<r_\mathrm{dec}<r_\mathrm{N}$ within the green region, which means that the shell continues to expand following the trajectory indicated by the arrow until reaching the triple coincidence $r_\mathrm{N}\sim r_\Delta\sim r_\mathrm{dec}$ at $\xi \sim 1$.

In the triangle region bound by $\xi \sim 1$ and $\xi \sim \sigma^{-1/6}$, $r_\mathrm{F}$ is the smallest radius.
If the shell starts the coasting phase here, it first forms a Newtonian RS at $r \sim r_\mathrm{F}$, and then begins to spread after passing $r_\mathrm{s}$.
Following the shell-spreading trajectory, the RS inevitably strengthens into a trans-relativistic shock while crossing and decelerating the shell at $\xi \sim 1$.

A common feature of this `thin shell' regime is that the shell-spreading effect ultimately leads to a trans-relativistic RS crossing and decelerating the entire shell at approximately the same time (the triple coincidence $r_\mathrm{N}\sim r_\Delta\sim r_\mathrm{dec}$ at $\xi \sim 1$).
The differences between the sub-regimes are also evident:
If the shell begins coasting in the triangle region ($1 \lesssim \xi_0 \lesssim \sigma_0^{-1/6}$), an RS is first formed before the shell spreads.
If the initial shell width is thinner ($\xi_0 \gtrsim \sigma_0^{-1/6}$), the shell spreading leads to RS formation.
Above the red dashed line ($\xi_0 \gtrsim \sigma_0^{-1/3}$), the RS is not even formed if shell spreading is omitted.
One plausible example is the numerical simulation by MGA09 with $\xi_0=0.5,1.1$ and $\sigma_0=1$.
Although their simulated results lie at the intersection of the regimes and thus are difficult to be categorized via order-of-magnitude arguments, the results support the claim that shell spreading (induced by rarefaction waves in MGA09) could lead to the formation of an otherwise impossible RS.

\subsection{`No RS' regime (red and white regions)}
\label{sec:nors}

The shell is directly decelerated by the FS without forming an RS, either from the beginning (red region) or after shell spreading (white region).
The radii associated with the RS are not relevant, as the RS is not formed ($r_\Delta < r_\mathrm{F}$).
This regime can be further divided into two sub-regimes according to whether the shell spreads before deceleration, as discussed below.

In the red region ($\sigma^{-1/3}\lesssim \xi \lesssim \sigma^{1/6}$), the deceleration radius becomes the smallest radius.
Consequently, as the shell propagates, it continues coasting until reaching the deceleration radius $r_\mathrm{dec}=r_\mathrm{dec,0}\sim l\gamma_0^{-2/3}(1+\sigma_0)^{-1/3}$, and begins to be decelerated by the FS following the Blandford-Mckee self-similar solution \citep{blandfordFluid1976PF}.

In the white region ($\sigma^{1/6}\lesssim \xi \lesssim \sigma$), the shell-spreading radius is the smallest radius.
Consequently, the coasting phase subsequently leads to shell spreading at $r\gtrsim r_\mathrm{s}$.
The relative magnitudes of the five radii are related to each other through $\xi$ and $\sigma$
according to relation~\eqref{eq:radiirelationspread}.
As the shell propagates to larger distances, both $\xi$ and $\sigma$ drop as $\xi\propto r^{-1/2},\sigma\propto r^{-1/2}$ (indicated by the arrow in the white region following $\xi \propto \sigma$; see Table~\ref{tab:1}).
Therefore, in the white region, the shell spreading does not lead to the formation of an RS as in the `thin shell' regime, but instead leads to deceleration.
The shell keeps expanding following the trajectory indicated by the arrow until $r\sim r_\mathrm{dec}$ at $\xi \sim \sigma^{1/6}$, where the shell is decelerated by the FS.

However, the boundary between the white region and the `thin shell' regime is solely determined by the estimate that a highly-magnetized shell spreads following $\xi \propto \sigma$.
This has not been discussed in the literature yet, and detailed numerical simulations are needed to determine the boundary between the `no RS' regime and the `thin shell' regime.
For this reason, we label the white region as the `no RS?' regime to acknowledge the ambiguity at the boundary.




\subsection{Comparison with previous works}
\label{sec:compare}

Here, we point out that Fig.~\ref{fig:2} is essentially the combination of Figure 4 in ZK05 and Figure 1 in GMA08, modified due to our current understanding of RS formation.
Figure 4 in ZK05 is the upside-down version of this diagram.
As done in our Fig.~\ref{fig:2}, both ZK05 and GMA08 correctly categorized the parameter space into three regimes, and interpreted the `thick shell' (region I in ZK05 and yellow region in Fig.~\ref{fig:2}) and `thin shell' (region III in ZK05 and green region in Fig.~\ref{fig:2}) regimes similar to our discussion.
However, there are two major improvements in our Fig.~\ref{fig:2} compared to ZK05 and GMA08:
\begin{itemize}
    \item[$\bullet$]
    The existence of a `no RS' regime is clarified.
    ZK05 suggested that RS may still form and even lead to broad emission peak in their region II (the red region in Fig.~\ref{fig:2}) based on the observation that $r_\Delta$ is the smallest radius.
    We find that this is not the case.
    This discrepancy arises because ZK05 did not include the RS forming radius in their calculations.
    Although the RS crossing radius $r_\Delta$ is the smallest radius in the red region, the RS does not even form since $r_\Delta < r_\mathrm{F}$.
    This further suggests the importance of including the RS forming radius $r_\mathrm{F}$ in the discussion of magnetized external shocks.
    This point is corrected in Figure 1 of GMA08, where they showed the RS forming regions similar to our diagram.
    \item[$\bullet$]
    The boundary between the `no RS' regime and the `thin shell' regime is updated, though some uncertainties remain.
    In ZK05, this boundary between their region II and III was found to be $\xi \sim \sigma^{1/3}$, where $r_\Delta \sim r_\mathrm{s}$.
    However, we have shown that the RS does not form in these regions, which makes $r_\Delta$ irrelevant to the evolution of the shell.
    In these regions, we should only discuss whether the shell spreads first or decelerates first.
    GMA08 proposed that the shell spreading only operates at low $\sigma$, which leads to an extended `no RS' regime as well as a reduced `thin shell' regime compared to our Fig.~\ref{fig:2}.
    They seem to indicate that when $\sigma \gtrsim 1/3$, the shell establishes causal contact through fast magnetosonic waves and limits shell spreading.
    We do not find this argument convincing, as there is no evidence that the variations in flow velocity can be suppressed by causal contact, although the causal contact may help maintain the coherence of magnetic field \citep[e.g.,][]{lyutikovGamma2003a}.
    In our Fig.~\ref{fig:2}, this boundary is put at $\xi \sim \sigma$, which is our estimate of whether the shell spreading leads to direct deceleration or RS formation.
    This is closely related to the shell-spreading process in the high-$\sigma$ regime, which has not been probed in previous works.
    The location of this boundary still merits further investigations, in particular through numerical simulations\footnote{
    The numerical simulations conducted by MGA09 did not explore this shell-spreading regime of the parameter space.
    Furthermore, we argue that the simulated results of MGA09 shown in Fig.~\ref{fig:1} cannot be plotted in Fig.~\ref{fig:2} for direct comparison (as was done in MGA09).
    This is because the parameters chosen in MGA09 for their Figure 4 are at the intersection of different regimes ($\xi_0,\sigma_0 \sim 1$), which may not be correctly categorized by an order-of-magnitude argument.
    They also adopted an artificially defined $\xi_\mathrm{eq}$ from a long-lasting ejecta to mimic the $\xi$ parameter evaluated at RS crossing for a finite-width shell, which may not be applicable for a direct comparison with Fig.~\ref{fig:2}.
    }.
\end{itemize}



\subsection{Estimating the critical degree of magnetization}
\label{sec:sigcr}

GRB parameters have wide ranges, e.g, initial Lorentz factor $\gamma_0 \sim 10^2-10^3$, initial shell width $\Delta_0 \sim 10^7-10^{13}$ cm, total energy $E \sim 10^{50}-10^{54}$ erg, and medium rest-frame number density $n_1 \sim 10^{-1}-10^3\ \mathrm{cm^{-3}}$
\citep{japeljPhenomenology2014A, gaoMorphological2015A, kumarphysics2015PRb, zhangPhysics2018TPoGBbBZI9CUP}.
Medium estimate gives
\begin{equation}
    \xi_0 \simeq 1.3 \times \gamma_{0,2.5}^{-4/3} \Delta_{0,11}^{-1/2} E_{52}^{1/6} n_{1,0}^{-1/6}\, ,
\end{equation}
where $Q_n \equiv Q/10^n$ is adopted in cgs units hereafter.
Note that $\xi_0$ is more sensitive to our choices of $\gamma_0$ and $\Delta_0$, and both `thin shell' ($\xi_0 \gtrsim 1$) and `thick shell' ($\xi_0 \lesssim 1$) are plausible scenarios.

As shown in Fig.~\ref{fig:2}, for the ejecta to eventually form an RS, the initial degree of magnetization should satisfy
\begin{equation}
    \sigma_0 \lesssim \sigma_\mathrm{0,cr} \simeq \max(\xi_0^{-3},\xi_0)\, ,
\end{equation}
which is sensitive to $\gamma_0$ and $\Delta_0$.
Here, $\sigma_\mathrm{0,cr}$ is the critical value of $\sigma_\mathrm{0}$ below which an RS can eventually form.
The most stringent constraint is achieved for $\xi_0\sim 1$.
Therefore, we find that all the ejecta with $\sigma_0 \lesssim 1$ should be able to eventually form an RS.
This is approximately the estimate given by GMA08, and is consistent with the argument that RS will always exist in the hydrodynamic cases.
Nevertheless, in the extreme cases where the shell is ultra-thin ($\xi_0 \sim 100$) or ultra-thick ($\xi_0 \sim 0.1$), even an initially highly-magnetized ejecta with $\sigma_0 \sim 10^3$ can eventually form an RS.

However, this is not to be confused with the magnetization of the shell when the RS forms, which we denote as $\sigma_\mathrm{F}$.
The RS forming condition gives $\sigma_\mathrm{F} < \sigma_\mathrm{F,cr} = (8/3) \zeta_\mathrm{F}$, where $\zeta_\mathrm{F}$ is the value of $\zeta$ parameter when the RS is formed.
Normally, it is difficult to estimate $\zeta_\mathrm{F}$, mainly because the number density $n_4$ at $r_\mathrm{F}$ depends on the dynamical history of the ejecta.
Here, we resort to Fig.~\ref{fig:2} to estimate $\sigma_\mathrm{F}$.
Based on scaling laws (see Table~\ref{tab:1}), we find
\begin{equation}
    \sigma_\mathrm{F}\simeq \left\{
    \begin{split}
        & \sigma_0,\ \xi_0\lesssim \min(\sigma_0^{-1/6},\sigma_0^{-1/3})\\
        & \sigma_0^{3/4}\xi_0^{-3/2},\ \xi_0^{-6}\lesssim\sigma_0\lesssim 1\\
        & \sigma_0^{3/2}\xi_0^{-3/2},\ 1\lesssim\sigma_0\lesssim \xi_0\, 
    \end{split}
    \right.
    \ .
\end{equation}
In the `thick shell' regime, $\sigma_\mathrm{F}$ has essentially the same value as $\sigma_0$, and thus is less constrained.
In principle, it could reach $\sim 10^3$ for ultra-thick shells, which is even greater than the estimate given by ZK05.
However, in the `thin shell' regime, only if the shell enters the triangle region ($1\lesssim \xi\lesssim \sigma^{-1/6}$) can an RS be formed, which means that $\sigma_\mathrm{F} \lesssim 1$.
Therefore, for the `thin shell' scenario, the degree of magnetization $\sigma$ typically drops to well below unity when the RS is formed \citep{granotImpulsive2011M, granotInteraction2012M, gaoPhotosphere2015A}, which then decreases further when the RS crosses and decelerates the shell as a trans-relativistic RS.

\section{Conclusions}
\label{sec:conclusions}

In this paper, we characterize the RS properties of magnetized relativistic outflows, and clear the confusion about the RS forming condition.
To this end, we extend the theoretical formulation of ZK05, and compare our theoretical predictions of RS properties with the simulated results of MGA09.
We revisit the RS forming conditions given by ZK05, GMA08, and \citet{lyutikovSimple2010PRE, lyutikovDynamics2011M}, and reconcile the differences between them.
Using the ZK05 condition to define a RS forming radius, we discuss the evolution of magnetized shell at the external shock stage.

Our conclusions are summarized as follows:
\begin{itemize}
    \item[$\bullet$]
    For the properties of the RS, we confirm previous arguments that a dynamically-important magnetic field ($\sigma \gtrsim 1$) suppresses the strength of RS \citep{zhangGamma2005Aa, gianniosexistence2008A, mimicaDeceleration2009A, granotInteraction2012M, zhangPhysics2018TPoGBbBZI9CUP}, forcing it to dissipate energy at efficiency less than $\sim 0.01-0.1$.
    \item[$\bullet$]
    For the RS forming condition, we argue that the ZK05 condition $\sigma < (8/3) \gamma_4^2 (n_1/n_4)$ is the correct condition, which can be exploited for the setup and consistency check of relevant numerical simulations \citep[e.g.,][]{aimechanical2021M, aiEnginefed2022ae}.
    We find the simulated results of MGA09 are more consistent with the ZK05 predictions \textit{rather than} the GMA08 condition.
    It is further shown that the GMA08 condition, though correct in concept, mistakenly gives a tighter constraint due to their procedure of order-of-magnitude estimate.
    If formulated more rigorously, we show in Appendix~\ref{sec:appendixb} that the physical motivation of GMA08 leads to the ZK05 condition.
    \item[$\bullet$]
    The evolution of a magnetized ejecta at the external-shock stage can be classified into three regimes: 
    (1) `thick shell' regime ($\xi_0 \lesssim 1, \sigma_0 \lesssim \xi_0^{-3}$) where the RS is formed and becomes relativistic, subsequently decelerating the shell while crossing it.
    (2) `thin shell' regime ($\xi_0 \gtrsim 1, \sigma_0 \lesssim \xi_0$) where a Newtonian RS is formed either after shell spreading or from the beginning, which then crosses and decelerates the shell as a trans-relativistic shock.
    (3) `no RS' regime ($\sigma_0^{-1/3} \lesssim \xi_0 \lesssim \sigma_0$) where the shell is decelerated by the FS, either after shell spreading or from the beginning, without forming an RS.
    Here, $(\xi_0,\sigma_0)$ refer to the values of ejecta parameters $(\xi,\sigma)$ at coasting phase.
    Therefore, all the ejecta with $\sigma_0\lesssim 1$ are capable of forming RSs, either initially or aided by shell spreading.
    The boundary between the `thin shell' regime and the `no RS' regime at high-$\sigma$ is still ambiguous, limited by our poor understanding of shell spreading in a highly magnetized ejecta.
    \item[$\bullet$]
    For a typical GRB problem, we can define the critical degree of magnetization $\sigma_\mathrm{F,cr}$ below which the RS is immediately formed.
    We find that this value is highly sensitive to the bulk Lorentz factor $\gamma_0$ and the shell width $\Delta_0$.
    For the RS formed in the `thin shell' scenario, the critical $\sigma_\mathrm{F,cr}$ is generally below unity, similar to the value proposed in GMA08.
    However, for the `thick shell' cases, the critical $\sigma_\mathrm{F,cr}$ is less constrained, which can potentially reach $100-1000$ for ultra-thick shells if there is a pre-existing coasting phase (similar to the value proposed in ZK05).
\end{itemize}

Here, we restrict our theoretical formulation to 1D outflows, with only some order-of-magnitude discussions on shell spreading.
Detailed numerical simulations are needed to probe the shell-spreading effect in magnetized outflows (as suggested in GMA08 and MGA09), as well as other instabilities potentially arising from multi-dimensional treatments \citep[as suggested in][]{lyutikovSimple2010PRE, lyutikovDynamics2011M}.

In addition, our discussions in Sec.~\ref{sec:rsradius} assume that there is always a coasting phase before shell spreading or external shocks.
This may not be true in the high-$\sigma$ regime \citep[see, e.g.][]{granotImpulsive2011M, granotInteraction2012M, gaoPhotosphere2015A}.
Whether the shell can reach a steady coasting phase depends on the acceleration mechanism, as well as the comparison between the coasting radius and other characteristic radii.
In the `thin shell' regime, the impulsive acceleration may accelerate the Poynting-flux-dominated ejecta beyond the fast magnetosonic point, leading to a prolonged, 
slower 
acceleration phase followed by deceleration 
\citep{granotImpulsive2011M, granotInteraction2012M}.
For thick shells that resemble long-lasting magnetized jets,
efficient
acceleration beyond the fast magnetosonic point is not trivial
\citep{beskinMHD1998M, tchekhovskoySimulations2008M, tchekhovskoyEfficiency2009A}, which may involve magnetic dissipation 
\citep{lyubarskyReconnection2001A, drenkhahnAcceleration2002A, drenkhahnEfficient2002A, kirkDissipation2003A} or external confinement 
\citep{komissarovMagnetic2009M, lyubarskyAsymptotic2009A}.
Unconfined steady-state jets in ideal MHD could still be accelerated efficiently if (a) the `rarefaction acceleration' operates at the onset of deconfinement \citep{komissarovRarefaction2010M, tchekhovskoyMagnetohydrodynamic2010NA}, or (b) the `causality surface' (where the flow loses lateral causal contact with the jet axis) is well beyond the fast magnetosonic point (where the causal contact is broken along the stream line) \citep{tchekhovskoyEfficiency2009A}.
However, the former mechanism may only be important shortly after the jet emerges from the stellar envelope \citep{komissarovRarefaction2010M, gaoPhotosphere2015A}, and the latter effect depends on the magnetic field configuration \citep{beskineffective2006M, tchekhovskoySimulations2008M, puProperties2020A, chenAnalytical2021A}.
These suggest that a coasting phase may be more plausible for thick shells.
Nevertheless, if analyzed more properly (e.g., by considering all the evolutionary stages of the ejecta), a strongly-magnetized shell could be directly decelerated or form an RS during the acceleration of the ejecta, which deserves further investigation.
The results shown here could be applied to such analyses, with further implications for the self-consistent lightcurve modelling of magnetized relativistic outflows.

\section*{Acknowledgements}

We thank the anonymous referee for helpful comments and Shunke Ai for helpful discussion on mechanical models.
J.-Z. Ma acknowledges the support from Tsien Excellence in Engineering Program (TEEP) in Tsinghua University.

\section*{Data Availability}

The code developed to perform the calculation in this paper is available upon request.
 



\bibliographystyle{mnras}
\bibliography{Shock} 




\appendix

\section{Jump conditions of magnetized relativistic shocks}
\label{sec:appendixa}

Following the symbol convention of ZK05, the upstream region (unshocked) is denoted as region 1 and the downstream region (shocked) is denoted as region 2, separated by the shock (denoted as `s').
$Q_{ij}$ ($Q_{a,ij}$) represents the quantity $Q$ ($Q_a$) in region $i$ in the rest frame of $j$ ($i = 1,2$ and $j = 1,2,s$).
Assuming a cold upstream, the shock jump conditions can be written in the most general forms as (see ZK05)
\begin{align}
    & \frac{e_2}{n_2 m_\mathrm{p} c^2} = (\gamma_{21}-1)f_{a,21}\, ,
    \label{eq:jc1}\\
    & \frac{n_2}{n_1} = \frac{\hat{\Gamma}_2 \gamma_{21} +1}{\hat{\Gamma}_2 - 1}f_{b,21}\, ,
    \label{eq:jc2}\\
    & \frac{p_{\mathrm{tot},2}}{p_2} = f_{c,21}\, ,
    \label{eq:jc3}
\end{align}
supplemented by the EOS with $p_2 = (\hat{\Gamma}_2 - 1) e_2$.
Here, $f_{a,21}, f_{b,21}, f_{c,21}$ are the correction factors contributed by the magnetic nature of the shock.
They are given by
\begin{align}
    & f_{a,21} = 1 - \frac{\gamma_{21}+1}{2u_{1s}u_{2s}}\sigma_1\, ,\\
    & f_{b,21} = \frac{\hat{\Gamma}_2 - 1}{\hat{\Gamma}_2 \gamma_{21} +1} \gamma_{21}\left(1+\frac{\beta_{21}}{\beta_{2s}}\right)\, ,
    \label{eq:fb}\\
    & f_{c,21} = 1 + \frac{u_{1s}}{2(\hat{\Gamma}_2-1)u_{2s}}\left(\frac{e_2}{n_2 m_\mathrm{p} c^2}\right)^{-1}\sigma_1\, .
\end{align}
The quantity $\beta_{2s}=\beta_{2s}(\sigma,\gamma_{21},\hat{\Gamma}_2)$ is found by solving a 
three order
equation \citep[see a detailed derivation in][]{maRelativistic2022M}
\begin{equation}
\begin{split}
    &\beta_{2s}(\beta_{2s}+\beta_{21})\left[\beta_{2s}-(\hat{\Gamma}_2-1)\frac{\beta_{21}}{1+\sqrt{1-\beta_{21}^2}}\right]\\
    &=\left[\beta_{2s}+\left(1-\frac{\hat{\Gamma}_2}{2}\right)\beta_{21}\right](1-\beta_{2s}^2)\sigma\, ,
\end{split}
\end{equation}
and $\beta_{1s}$ is found through the Lorentz transformation $\gamma_{1s} = \gamma_{21}\gamma_{2s}(1+\beta_{2s}\beta_{21})$.
Combine equations~\eqref{eq:jc1}-\eqref{eq:jc3} with the EOS, and we get the downstream total pressure
\begin{equation}
    p_{\mathrm{tot},2} = (\hat{\Gamma}_2 \gamma_{21} +1)(\gamma_{21}-1)n_1 m_\mathrm{p} c^2 F_{21}\, ,
\end{equation}
where $F = f_a f_b f_c$.

\section{A rigorous derivation of the contact radius and the RS crossing radius}
\label{sec:appendixb}

\begin{figure}
	\includegraphics[width=\columnwidth]{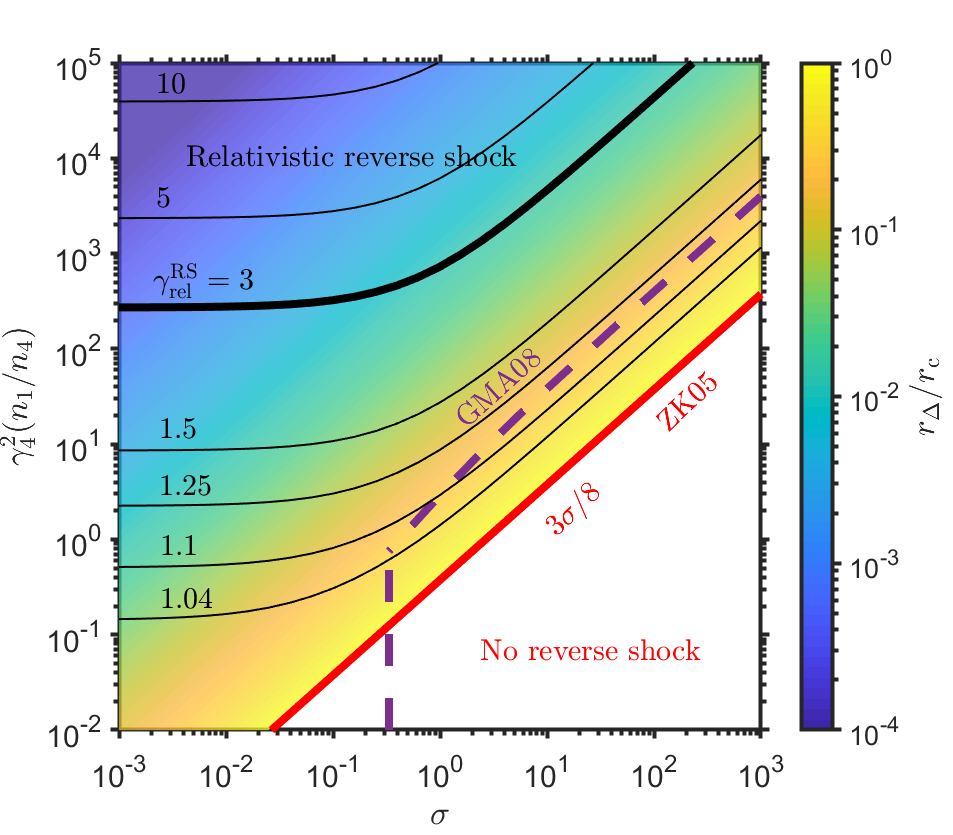}
    \caption{
    Ratio of the RS crossing radius $r_\Delta$ to the contact radius $r_\mathrm{c}$ in the parameter space.
    The line styles are associated with the RS relative Lorentz factor and different treatments of RS forming conditions, as in Fig.~\ref{fig:1}.
    The RS crossing radius coincides with the contact radius at the ZK05 condition (red solid line) \textit{rather than} the GMA08 condition (purple dashed line).
    }
    \label{fig:a1}
\end{figure}

Following GMA08, the contact radius $r_\mathrm{c}$ is defined as the radius where the fast magnetosonic wave crosses the shell at the expansion timescale, i.e.,
\begin{equation}
    r_\mathrm{c} \equiv \frac{\Delta}{\beta_4 - \beta_\mathrm{w}}\, ,
\end{equation}
where $\beta_\mathrm{w}$ is the dimensionless wave speed in lab frame given by the Lorentz transformation $\beta_\mathrm{w}=(\beta_4-\beta_\mathrm{A})/(1-\beta_4 \beta_\mathrm{A})$.
Here, $\beta_\mathrm{A} = \sqrt{\sigma/(1+\sigma)}$ is the dimensionless Alfven speed.
The contact radius can thus be written rigorously as
\begin{equation}
    r_\mathrm{c} = \Delta \gamma_4^2  \left(\sqrt{\frac{1+\sigma}{\sigma}} - \beta_4\right)\, ,
    \label{eq:rcrigorous}
\end{equation}
For typical GRB cases with relativistic ejecta, it reduces to the expression~\eqref{eq:rc}
\begin{equation}
    r_\mathrm{c} \simeq \Delta \gamma_4^2  \left(\sqrt{\frac{1+\sigma}{\sigma}} - 1\right)\, .
\end{equation}

Following \citet{sariHydrodynamic1995A} and ZK05, the RS crossing radius is defined as the radius where the RS crosses the shell \citep[for a rigorous derivation see, e.g., Chapter 8.7.1 of][]{zhangPhysics2018TPoGBbBZI9CUP}, i.e.,
\begin{equation}
    r_\Delta \equiv \frac{\Delta}{\beta_4 - \beta_3}\left(1-\frac{\gamma_4 n_4}{\gamma_3 n_3}\right)\, ,
    \label{eq:rrsdef}
\end{equation}
where the last factor is taken into account due to the compressed shell after RS crossing.
Substituting the shock jump condition~\eqref{eq:jc2}\eqref{eq:fb} and Lorentz transformation $\gamma_{43} = \gamma_4\gamma_3(1-\beta_4\beta_3)$ into equation~\eqref{eq:rrsdef}, we find
\begin{equation}
\begin{split}
    r_\Delta 
    & = \frac{\Delta}{\beta_4 - \beta_3}\left(1-\frac{\beta_{4s}(1-\beta_3^2)}{ (1-\beta_4\beta_3)(\beta_{4s}+\beta_{43})}\right)\\
    & = \Delta\, \frac{\beta_{4s}\beta_3(\beta_3-\beta_4) + (1-\beta_4\beta_3)\beta_{43}}{(\beta_4-\beta_3)(1-\beta_4\beta_3)(\beta_{4s}+\beta_{43})}\\
    & = \Delta \frac{1-\beta_{4s}\beta_3}{(1-\beta_4\beta_3)(\beta_{43}+\beta_{4s})}\, .
\end{split}
\end{equation}
As the RS grows weak and approaches the point of vanishing, we get $\beta_{4s} \rightarrow \beta_\mathrm{A}, \beta_{4s} \rightarrow \beta_4, \beta_{43} \rightarrow 0$.
Consequently, at the verge of RS formation, the RS crossing radius approaches the contact radius, i.e., $r_\Delta \rightarrow r_\mathrm{c}$.
For a system with an FS that is at least mildly relativistic, we have $\beta_3 \simeq \beta_4 \sim 1$.
We thus obtain
\begin{equation}
    r_\Delta \simeq \Delta \gamma_4^2 \frac{1-\beta_{4s}}{\beta_{43} + \beta_{4s}}\, ,
    \label{eq:rrsrigorous}
\end{equation}
which gives
\begin{equation}
    \frac{r_\Delta}{r_\mathrm{c}} \simeq \frac{1-\beta_{4s}}{(\beta_{43} + \beta_{4s})\left(\sqrt{(1+\sigma)/\sigma}-1\right)}\, .
    \label{eq:radiiratio}
\end{equation}
With the magnetization degree $\sigma$ fixed, $\beta_{4s}$ drops as $\beta_{43}$ decreases.
The ratio of the two radii thus monotonically increases as the RS grows weak until the two radii become equal to each other at the pressure balance condition~\eqref{eq:ZK05} suggested by ZK05.

In Fig.~\ref{fig:a1}, we present the ratio~\eqref{eq:radiiratio} of two radii in the parameter space, with GMA08 condition plotted as the purple dashed line and ZK05 condition as the red solid line.
As the RS becomes weak, the RS crossing radius increases until it coincides with the contact radius at the ZK05 condition \textit{rather than} the GMA08 condition.
Here, we performed a more careful derivation for the characteristic radii to show that the physical motivation of GMA08 condition is fully consistent with ZK05.
However, due to the order-of-magnitude estimates in GMA08, the condition given in GMA08 is mistakenly tighter than the actual RS forming condition given in ZK05.

\section{Scaling laws for the dynamical evolution of a magnetized shell}

\begin{table*}
	\centering
	\caption{
	Scaling laws for the dynamical evolution of a magnetized shell.
	}
	\begin{threeparttable}
	\begin{tabular}{ccccccc}
	    \hline
	    \multicolumn{7}{c}{General assumptions\tnote{a}$\ $ : $E=4\pi m_\mathrm{p} c^2 \gamma_4^2 n_4 r^2 \Delta (1+\sigma) = \mathrm{const}, \sigma\propto n_4 r^2$}\\
		\hline
		\multicolumn{2}{c}{Evolutionary stage}
		& Feature & $\log\xi/\log r$ & $\log n_4/\log r$ & $\log\sigma/\log r$  & $\log\xi/\log\sigma$
		\\
		\hline
		\multicolumn{2}{c}{Coasting} & $\gamma_4\simeq \mathrm{const},\Delta\simeq\mathrm{const}$ & $0$ & $-2$ & $0$ & $-$
		\\
		\hline
		\multirow{2}{*}{Deceleration by RRS, $r_\mathrm{N} \lesssim r \lesssim r_\Delta$} & $\sigma \ll 1$ & \multirow{2}{*}{$\gamma_4 \propto r^{-1/2},\Delta \simeq \mathrm{const}$\tnote{b}} & \multirow{2}{*}{$2/3$} & $-1$ & $1$ & $2/3$ 
		\\
		& $\sigma \gg 1$ & & & $-3/2$ & $1/2$\tnote{d} & $4/3$ 
		\\
		\hline
		\multirow{2}{*}{Shell spreading, $r_\mathrm{s} \lesssim r \lesssim r_\mathrm{dec}$} & $\sigma \ll 1$ & \multirow{2}{*}{$\Delta \propto r, \gamma_4 \simeq \mathrm{const}$\tnote{c}} & \multirow{2}{*}{$-1/2$} & $-3$ & $-1$\tnote{d} & $1/2$ 
		\\
		& $\sigma \gg 1$ & & & $-5/2$ & $-1/2$ & $1$ 
		\\
		\hline
	\end{tabular}
	\begin{tablenotes}[para]\footnotesize
        \item[a] \citet{mimicaDeceleration2009A}.
        \item[b] \citet{kobayashiHydrodynamics1999A}.
        \item[c] \citet{meszarosGasdynamics1993A, piranHydrodynamics1993M}.
        \item[d] \citet{granotInteraction2012M}.
    \end{tablenotes}
	\end{threeparttable}
	\label{tab:1}
\end{table*}

In Table~\ref{tab:1}, we present the scaling laws of shell parameters as functions of radius in different evolutionary stages.
It should be pointed out here that the scaling laws in high-$\sigma$ regime are uncertain.
This is mainly because the evolution of $\gamma_4$ and $\Delta$ as the shell travels to larger distances are less clear for $\sigma \gg 1$.
In particular, a coasting phase may not exist in high-$\sigma$ regime \citep{granotImpulsive2011M, granotInteraction2012M, gaoPhotosphere2015A}, which would make some of our discussions irrelevant to the real scenario.
Here, we assume that the total energy of the ejecta is conserved during its evolution, and $\sigma \propto n_4 r^2$ as pointed out by \citet{mimicaDeceleration2009A}.
We extrapolate the scaling laws of $\gamma_4(r)$ and $\Delta(r)$ for $\sigma \ll 1$ to high-$\sigma$ regime.
Although some results given here seem to agree with previous analytical estimates for magnetized shells \citep{granotInteraction2012M}, more rigorous analyses are needed.

\bsp	
\label{lastpage}
\end{document}